\documentclass[aps,prd,twocolumn,nofootinbib,showpacs,preprintnumbers]{revtex4}

\usepackage{epsfig}
\usepackage{amsmath}
\usepackage{amssymb}
\usepackage[dvipdfm]{hyperref}

\def\e{\,{\rm e}\,}
\def\d{{\rm d}}
\def\i{{\rm i}}
\def\D{{\cal D}}
\def\ym{y_{\rm min}}
\def\del{\varepsilon}

\newcommand{\rf}[1]{(\ref{#1})}
\newcommand{\eq}[1]{Eq.~(\ref{#1})}
\def\be{\begin{equation}}
\def\ee{\end{equation}}
\def\bea{\begin{eqnarray}}
\def\eea{\end{eqnarray}}

\newcommand{\non}{\nonumber \\*}
\newcommand{\ie}{{i.e.}\ }
\def\la{\lesssim}
\def\ga{\gtrsim}

\begin{document}

\preprint{ITEP--TH--57/11}

\title{Remarks on Holographic Wilson Loops and the Schwinger Effect} 

\author{Jan Ambj\o rn}

\affiliation{The Niels Bohr Institute\\
\mbox{Blegdamsvej 17, 2100 Copenhagen \O, Denmark}}
\email{ambjorn@nbi.dk}

\author{Yuri Makeenko}

\affiliation{Institute of Theoretical and Experimental Physics\\
B.~Cheremushkinskaya 25, 117218 Moscow, Russia}
\email{makeenko@itep.ru} 

\date{January 5, 2012}

\begin{abstract}
We extend Douglas' solution of the problem of finding minimal surfaces
to anti-de Sitter space. The case of a circle as a boundary contour is
elaborated. We discuss applications to ${\cal N}=4$ super Yang--Mills:  
a circular Wilson loop and the Schwinger process, where we
calculate the $1/\sqrt{\lambda}$ correction to the critical value
of constant electric field. 
\end{abstract}



\maketitle 

\section{Introduction}

The Schwinger effect~\cite{Sch51} of pair production in a constant 
electric field is one of the beautiful predictions of QED. 
The production rate of a pair of particles with masses $m$
and charges $\pm e$ is exponentially suppressed for weak fields $E$ as
\be
P\propto \e^{-\pi m^2/|e E|},
\label{dr}
\ee
where the exponent has the meaning of a classical Euclidean action associated 
with the tunneling. With increasing $E$, fluctuations 
about the classical (Euclidean) 
trajectory, which has the form of a circle 
of radius $R=m/|e E|$, become more and more important, but
nothing special happens even for $|e E|\ga  1/m^2$, when the saddle-point
approximation~\cite{AAM82} in the path integral over 
(pseudo)particle trajectories ceases to be applicable. 

This smooth behavior drastically differs from that \cite{FTs85,Bur86,BP92}%
\footnote{For a review see Ref.~\cite{AMSS00}.} in string theory,
where there exists an instability for the fields larger than a certain critical
value of the order of the string tension: $|e E_c|\sim 1/2\pi\alpha'$.
This instability is apparently not related to the Schwinger effect and
takes place even for a neutral string with opposite charges at the ends,
thus occurring because stretching of the string then costs negative energy.

Recently, a very interesting conjecture about 
an existence of such a critical electric field  for ${\cal N}=4 $ 
super Yang-Mills (SYM) has been made in Ref.~\cite{SZ11}, 
based on a holographic description~\cite{GSS02} of the Schwinger effect via the
AdS/CFT correspondence. In this approach the saddle-point trajectory
is governed in the supergravity approximation by a minimal surface
spanned by a circle. The goal of the present article 
is to account for
fluctuations about this minimal surface in anti-de Sitter (AdS), 
which result in
a preexponential factor. We evaluate the decay rate using a
representation of the Wilson loop in ${\cal N}=4$ SYM through a path
integral over reparametrizations of the boundary circle with the 
action prescribed by AdS/CFT, that holographically captures 
fluctuations in the bulk. We show that 
the fluctuations do not cure the instability, and the critical value 
of electric field is simply shifted in the quadratic approximation
(as is displayed in \eq{Ec} below).
Our results confirm the expectation that the Schwinger effect in 
${\cal N}=4$ SYM at strong coupling
does not look as it does in QED but is rather as it would appear 
in string theory.

\section{The setup}

The saddle-point (Euclidean) action that determines the exponent
of the production rate in a constant electric field 
is given by the minimum of 
\be
S= 2\pi R m -\pi |e E | R^2 - \ln W\left({\rm circle}\right)
\label{Seff}
\ee
with respect to the radius $R$ of the circle.
This effective action emerges after performing the path integral
over (pseudo)particle trajectories, representing the vacuum-to-vacuum
amplitude in an external constant electric field.
In the path integral, first it is shown that the integral over the proper time
has a saddle point, and then one can show that 
the saddle-point trajectory is a circle
with (a large) radius $R=m/|e E|$~\cite{AAM82}. The circle lies in
the $\mu,\nu$-plane, when the constant electric  field $E$ is 
given by the $\mu,\nu$-component of the field strenght $F_{\mu\nu}$.
The existence of this saddle point is justified for small $ |e E|$, 
when the logarithm of the Wilson
loop on the right-hand side of \eq{Seff}
is subleading at weak couplings and contributes only to the preexponential. 

The holographic description of the Schwinger effect in SYM
relies~\cite{GSS02} on the spherical solution~\cite{BCFM98,DGO99} 
of the Euler--Lagrange equations
for the minimal surface in $AdS$ enclosed by a circle in the boundary.
We shall write it for the upper half-plane (UHP) parametrization
of the surface: $z=x+\i y$ ($y>0$), which is customary in string theory,
using the standard embedding space coordinates $Y_{-1}$, $Y_0$, $Y_1$,
$Y_2$, $Y_3$, $Y_4$ obeying
\be
Y\cdot Y \equiv -Y_{-1}^2- Y_0^2+Y_1^2+Y_2^2+Y_3^2+Y_4^2=-1.
\label{=1}
\ee 
The solution reads
\bea
Y_1&= &\frac{1-x^2-y^2}{2y},\quad
Y_2 = \frac{x}{y}, \non
Y_{-1} &= &\frac{1+x^2+y^2}{2y},\quad
Y_4 = Y_0=Y_3=0,
\label{solu}
\eea
or
\begin{subequations}
\bea
Z&\equiv&\frac{R}{Y_{-1}-Y_4}=R\frac{2y}{1+x^2+y^2} ,\\
X_1&\equiv& Z Y_1=R\frac{1-x^2-y^2}{1+x^2+y^2},\\
X_2&\equiv& Z Y_2=R\frac{2x}{1+x^2+y^2},
\eea
\label{PP}
\end{subequations}
on the Poincare patch, so the induced metric
\be
\d \ell^2 = \frac{\d x^2+\d y^2}{y^2}.
\label{Poin}
\ee 
is the Poincare metric of the Lobachevsky plane.
The solution~\rf{PP} obeys $X_1^2+X_2^2+Z^2=R^2$ and
corresponds to a circle of the radius $R$ in the boundary when $Z=0$.

For these coordinates the Euler--Lagrange equations
in the embedding $Y$-space are
\be
\left(\Delta-2\right)Y_i=0,\quad
\Delta =y^2\left(
\frac{\partial^2}{\partial x^2}+\frac{\partial^2}{\partial y^2}
\right)
\label{eqY}
\ee
and the ``mass'' 2   
arises because of the presence of the Lagrange multiplier which is used
to implement \eq{=1}.


\section{Dirichlet Green function and Poisson formula in $\mathbf{AdS}$}

As in flat space, we found it most convenient to use
an extension of Douglas' algorithm~\cite{Dou31} for finding
minimal surfaces to the Lobachevsky plane. 
Our program is to first  construct the Dirichlet Green function of \eq{eqY}
on the Lobachevsky plane, and then to derive the version of the Poisson formula
relevant to  the Lobachevsky plane. This formula 
will then allow us to reconstruct the minimal surface 
from its boundary value, so
the problem of finding the minimal surface will be reduced to the
problem of minimizing a boundary functional with respect to
reparametrizations. Finally, we use this boundary functional for
evaluations of bulk fluctuations about the minimal surface.

The Dirichlet Green function on the
Lobachevsky plane is a function of the (geodesic) distance between 
the images of the points
$(x_1,y_1)$ and $(x_2,y_2)$, which is determined by the metric \rf{Poin} 
to be
\be
L^2=\frac{(x_1-x_2)^2+(y_1-y_2)^2}{4 y_1 y_2}.
\ee
Acting by the operator on the left-hand side of \eq{eqY},
we obtain the Legendre equation whose solution for
the Dirichlet Green function is
\bea
G\left(x_1,y_1;x_2,y_2\right)&=&
-\frac{3}{4\pi} \Big(\frac{(x_1 - x_2)^2 + y_1^2 + 
      y_2^2}{4 y_1 y_2} \non &&\times \ln
\frac{(x_1 - x_2)^2 + (y_1 - y_2)^2}{(x_1 - x_2)^2 + (y_1 + y_2)^2}
+1 \Big).
\non &&
\label{Green}
\eea

To obtain the Poisson formula, which reconstructs a harmonic function
in the Lobachevsky plane (\ie a function which obeys \eq{eqY}) 
from its value at the boundary, we take the normal derivative 
of \eq{Green} near the boundary
at a certain minimal value $y_2=\ym$ to which the boundary is moved as
usual to regularize divergences:
\bea
\left.
\frac{\partial G\left(x_1,y_1;x_2,y_2\right)}{\partial y_2}\right|_{y_2=\ym}
&=&
\frac{2y_1^2 \ym}{\pi((x_1-x_2)^2+y_1^2)^2}\non
&&+{\cal O}(\ym^3).
\eea
We shall return soon to a physical meaning of this procedure.
Finally, we obtain
\be
Y_i(x,y)= \int_{-\infty} ^{+\infty}\frac {\d s} \pi
\, \frac{2Y_i(t(s)) y^2 \ym}{((x-s)^2+y^2)^2},
\label{LDir}
\ee
where $Y_i(t(s))$ is the boundary value and the function $t(s)$ is 
a possible reparametrization of the boundary, which plays a crucial
role in Douglas' algorithm.
This is an extension of the well-known Poisson formula
to the Lobachevsky plane.

It is instructive to see how the known solution~\rf{solu} for a circular 
boundary is reproduced by \eq{LDir} from the boundary values
\bea
Y_1(t) &= &\frac{1-t^2}{2\ym},\quad
Y_2(t) =  \frac{t}{\ym},\non
Y_{-1}(t) &= &\frac{1+t^2}{2\ym},\quad
Y_0(t) = Y_3(t) =Y_4(t)  =  0 ~~~
\label{63}
\eea
for $t(s)=s$, which means that no reparametrization of the boundary is required
for a circle, in analogy with the situation for the  ordinary Euclidean plane. 
The reason for this is that
the coordinates in use are conformal for a circle. 
Note that $\ym$ is nicely canceled, when \rf{63} in substituted in
\eq{LDir}. 

\section{An extension of Douglas' functional to $\mathbf{AdS}$}

As in flat space,  to obtain the minimal surface
we have to minimize the quadratic action, which now reads
\bea
S&=&\int \d x \,\d y 
\left[\frac 12 \partial_a Y(x,y) \cdot \partial _a Y(x,y)\right.\non
&&\left.+ \frac{\xi}{y^2}\left( Y(x,y)\cdot Y(x,y)+1 \right)\right],
\label{Squa}
\eea
where $Y_i(x,y)$ is recovered in UHP from the boundary value~\rf{63} 
by \eq{LDir} and the Lagrange multiplier $\xi(x,y)=1$ at the minimum.
This obtained value of $S$ has to be minimized with 
respect to the functions $t(s)$, reparametrizing the boundary.
The minimization is required for $Y_i$'s to obey
 a conformal gauge, where $\sqrt{g}$ would coincide with
the quadratic integrand in \eq{Squa}.
Remarkably, this can be formulated as the problem 
of minimizing a boundary functional which is an extension
of the flat-space Douglas integral 
\be
S_{\rm flat}=\frac{1}{4\pi} \int {\d s_1}
\int {\d s_2}\,\frac{
(x_B(t(s_1))- x_B(t(s_2)))^2 }{ (s_1-s_2)^2}
\label{oDou}
\ee
to AdS space. 

The Douglas integral~\rf{oDou} turned out to be extremely useful for
representing the area-law behavior of large Wilson loops in QCD.
Reference~\cite{MO08} contains a detailed description of this method.
An advantage of using such a representation of the minimal area is that
path integrals over trajectories $x^\mu(t)$ are now Gaussian and
easily doable, while nonlinearities are encoded in a
path integral over reparametrizations, whose extension to ${\cal N}=4$
will be soon considered.

Because $Y_i$'s obey \eq{eqY}, the integral over $y$ in \eq{Squa} 
reduces to a boundary term, after which the integral over $x$ yields
\begin{widetext}
\be
S= -\frac{1}\pi  \int {\d s_1}
\int {\d s_2}\,
Y_B(t(s_1))\cdot Y_B(t(s_2))\,\ym^2\left[\frac{1}{(s_1-s_2)^4}\right]_{\rm reg}
\label{ADou}
\ee
with
\be
\left[\frac{1}{(s_1-s_2)^4}  \right]_{\rm reg}=
\left( 
\frac{1}{((s_1-s_2)^2+4\ym^2)^2}
 +\frac{32\ym^2}{((s_1-s_2)^2+4\ym^2)^3}
 -\frac{384 \ym^4}{((s_1-s_2)^2+4\ym^2)^4}  \right).
\label{Greg}
\ee
\end{widetext}
This is the required boundary functional whose minimum with respect to
the functions $t(s)$ equals the minimal area.

The integral on the right-hand side of \eq{ADou} looks pretty similar
to that in \eq{oDou}, while the denominator in \eq{ADou} is $(s_1-s_2)$ to 
degree four rather than square as in \eq{oDou}. This is a manifestation
of the well-known divergences which are regularized by shifting the
boundary from $y=0$ to $y=\ym$. In the dual language of D-branes this
corresponds~\cite{Mal98,RY} to the breaking of the $U(N)$ symmetry down to 
$U(N-1)\times U(1)$ by assigning a finite mass to the $U(1)$ gauge boson.
If this mass is associated with shifting the boundary to the slice $Z=\del$, 
then 
\be
\ym(t)=\frac\del{2R}(t^2+1) 
\label{ymvsrm}
\ee
from \eq{63}.

The right-hand side of \eq{ADou} always diverges like 
\be
S_{\rm div}= 2\pi \frac {R-\del}\del,
\ee
which
comes from the domain $(s_1-s_2)\sim \ym$. It is universal and does not 
depend on the reparametrizing function $t(s)$. Subtracting the divergent
part, we obtain for the regularized part
\bea
\lefteqn{
S_{\rm reg}\equiv S-S_{\rm div}
=\frac{1}{2\pi}  \int {\d s_1}
\int {\d s_2} }\non
&&\times(Y_B(t(s_1))- Y_B(t(s_2)))^2\ym^2
\left[\frac{1}{(s_1-s_2)^4}\right]_{\rm reg}~~.
\label{Sreg}
\eea
The domain $(s_1-s_2)\sim \ym$ now gives a finite contribution to
this integral in view of the important formula
\be
\int {\d s}\,
s^2 \left[\frac{1}{s^4}  \right]_{\rm reg}=0.
\label{impo}
\ee

\section{Reparametrization path integral in ${\cal N}=4$ SYM}

We represent the circular Wilson loop in ${\cal N}=4$ SYM by
the  reparametrization path integral of the form
\be
W\left(\hbox{circle}\right)= \e^{-\sqrt{\lambda}S_{\rm div}/2\pi} 
\int \D_{\rm diff} t(s)
\e^{-\sqrt{\lambda}S_{\rm reg}[t]/2\pi},
\label{ansatz}
\ee
where
\be
S_{\rm reg}[t]=\frac1{2\pi}
\int \d s_1 \d s_2\,{\left(t(s_1)-t(s_2)\right)^2}
\left[\frac{1}{(s_1-s_2)^4}  \right]_{\rm reg}
\label{subt}
\ee
since $S_{\rm div}$ does not depend on the reparametrization as is
already pointed out.
The constant $\sqrt{\lambda}$ is prescribed by the AdS/CFT correspondence
to be
\be
\sqrt{\lambda}= \frac{R^2_{AdS}}{\alpha'},
\ee
but we shall simply consider it as a parameter of the ansatz to be fixed
by comparing with the Wilson loop in the ${\cal N}=4$ SYM perturbation theory.

Let us substitute for the reparametrizing function
\be
t(s)=s+\frac{1}{\sqrt[4]{\lambda}} \,\beta(s).
\ee
Because of \eq{impo} we then have
\be 
\sqrt{\lambda}S_{\rm reg}=  
\frac{1}{2\pi}\int \d s_1 \d s_2\left(\beta(s_1)-\beta(s_2)\right)^2
\left[\frac{1}{(s_1-s_2)^4}  \right]_{\rm reg}.
\label{integ}
\ee
While \eq{integ} is exactly equivalent to \eq{subt}, we shall restrict
ourselves by
an expansion in $1/{\sqrt[4]\lambda}$ to quadratic order
because the measure in the path integral~\rf{ansatz} is the one
for integrating over subordinated functions with $\d t(s)/\d s \geq 0$
and, as explicitly constructed in Ref.~\cite{MO08}, 
is highly nonlinear. Only to the quadratic order it can be substituted by
the ordinary Lebesgue measure.

Before evaluating the path integral~\rf{ansatz}, it is worth noting that 
the integral~\rf{integ} has three zero modes
\be
\beta_1(s)=1,\quad
\beta_2(s)=s,\quad   \beta_3(s)=s^2, 
\label{sl2t}
\ee 
which is a consequence of three $SL(2,\Bbb{R})$ symmetries.
For the second and third ones, \eq{impo} is again important. 

These three zero modes result in a preexponential factor of $\lambda^{-3/4}$
in a full analogy with the string theory analysis~\cite{DG00}.
We thus obtain from the ansatz~\rf{ansatz} at large $\lambda$:
\be
W\left(\hbox{circle}\right)\propto \lambda^{-3/4}\e^{\sqrt{\lambda}},
\label{Wfin}
\ee
reproducing the result~\cite{ESZ00} for the ${\cal N}=4$ SYM
perturbation theory, providing $\lambda$ is identified with the
't~Hooft coupling. Since fermions and the RR field, which are 
present in the IIB string representation of the ${\cal N}=4$ SYM, will manifest
themselves only to next orders, we believe that the constant factor
in \eq{Wfin} is also calculable like that~\cite{KT08}
in the string representation.

\section{Reparametrization path integral in ${\cal N}=4$ SYM (continued)}

In the derivation of \eq{Wfin}, we have mostly paid attention to
the dependence of the result on $\lambda$ rather than on $1/\del$
which plays the role of the $U(1)$ boson mass~\cite{Mal98,RY}
\be
m=\frac{\sqrt{\lambda}}{2\pi \del}
\label{Wmass}
\ee 
as is already mentioned.
We shall now concentrate on the dependence of $W\left(\hbox{circle}\right)$
on $\del$,  looking in detail at the divergences regularized by $\del$.
We are thus interested in the contributions 
from the reparametrization path integral to the effective action,
which are important  at small $\del$.

The calculation is pretty much similar to that of 
Ref.~\cite{MO10a} for a $T\times R$ rectangle in flat space,
where the L\"uscher term was obtained from the path integral over
reparametrizations.
In that case $T/R$ was large, now $R/\del$ is large. 
The idea is to perform a mode expansion
\be
\beta(s)= \sum _n \beta_n f_n(s)
\label{modes}
\ee
using a complete set of orthogonal basis functions 
$f_n(s)$ (in general complex ones obeying $f_{-n}(s)=f^*_n(s)$),
and then do the Gaussian integrals over $\beta_n$'s.
We can restrict ourselves by those modes for which the integral~\rf{integ}
has maximal ``divergence'' $\sim (R/\del)^\nu$. 
We then obtain
\be
\prod_n \left(\frac R{\del}\right)^{-\nu/2} = 
\left(\frac{R}{\del}\right)^{\nu/2}
=\e^{\frac \nu 2 \ln (R/\del)},
\label{QQ}
\ee
where the product goes over those modes for which the integral~\rf{integ}
is $\sim (R/\del)^\nu$. 
We have used here the $\zeta$-function regularization of the product
and taken into account that  $f_n(s)$'s are complex functions, so $n$
ranges from $-\infty$ to $+\infty$. 

What is the value of $\nu$? We have no reason to expect 
that typical functions in the path
integral over $\beta(s)$ are continuous, as it is the case for usual path
integrals with Wiener measure. Moreover, for smooth functions
we can substitute 
\mbox{$(\beta(s_1)-\beta(s_2))^2=$} \mbox{$(s_1-s_2)^2 (\d \beta(s_1)/\d s_1)^2$}
and their contribution to~\rf{integ} vanishes in view of \eq{impo}.
This is intimately linked to the above mentioned $SL(2,\Bbb{R})$ symmetry 
of the integral. In general, $\nu$ is determined by the Hausdorff 
dimension of $\beta(s)$. We assume that typical trajectories
in the reparametrization path integral have Hausdorff dimension zero%
\footnote{We remind that the Hausdorff dimension of the usual Brownian 
trajectories is one half.},
as was advocated in Ref.~\cite{BM09}. This corresponds to $\nu=3$.
Some more arguments in favor of this are given in Appendix~\ref{appA},
where we discuss in detail the Fourier expansion of $\beta(s)$.

\section{Schwinger effect in ${\cal N}=4$ SYM}

In the gravity approximation, when fluctuations about the minimal surface are
not taken into account, the action~\rf{Seff} reads~\cite{GSS02}
\be
\sqrt{\lambda} S_{\rm cl}= \sqrt{\lambda} \pi\left( \cosh\rho -1
-\frac{|e E|}{m^2}\sinh^2 \rho \right),
\label{Scl}
\ee
where $\sinh\rho = R/\del=2\pi m R /\sqrt{\lambda}$.
This formula is applicable, strictly speaking, for  
$|e E|\la m^2$, when
the minimization of $S_{\rm cl}$ with respect to $\rho$ gives
\be
\cosh\rho_0 = \frac{2\pi m^2}{|e E|\sqrt{\lambda}}.
\label{solu0}
\ee
As was pointed out in Ref.~\cite{SZ11}, this equation has no solution
for $\rho_0$ when $|e E|>2 \pi m^2/\sqrt{\lambda}$, 
which implies the existence of a critical electric field.

We are now in a position to answer the question as to 
how fluctuations about the minimal surface 
affect this very interesting result. 
The calculation of their contribution to the effective
action has been already obtained in \eq{QQ}.
For the sum of $S_{\rm cl}$
plus the contribution from fluctuations about
the minimal surface in the quadratic approximation we have
\bea
\sqrt{\lambda} S_{\rm cl+1loop}&=& \sqrt{\lambda} \pi\left( \cosh\rho -1
-\frac{|e E|}{m^2}\sinh^2\rho \right) \non
&&-\frac \nu2 \ln \cosh\rho.
\label{Scl1}
\eea
The negative sign for the contribution from
the fluctuations in the second line of this formula is like
for the L\"uscher term in string theory. We have mentioned already this
analogy, but would like to emphasize that it may have far-reaching consequences.

The minimum of the effective action~\rf{Scl1} is now reached for
\be
\frac{1}{\cosh\rho_0 }= \frac{\sqrt{\lambda}}\nu
\left(1-\sqrt{1-\frac{\nu |e E|}{\pi m^2}}\right),
\label{solu1}
\ee
so the solution \rf{solu0} is only slightly modified by
the quantum fluctuations. They simply shift the
critical value of the constant electric field to the value  
\be
|e E_c| = \pi m^2 \left(  \frac{2}{\sqrt{\lambda}}-\frac{\nu}{\lambda} \right),
\label{Ec}
\ee
where $\nu=3$ as is argued. 
Thus the quantum fluctuations about the minimal surface result
in a $1/\sqrt{\lambda}$ correction at large $\lambda$, as it might be expected. 

Our final comment is on how the one-loop effective action~\rf{Scl1}
agrees with that resulting in superstring theory  
from semiclassical fluctuations about the minimal surface. 
The case of an open superstring in $AdS_5\times S^5$ with the ends lying in
the boundary circle was elaborated in Refs.~\cite{DGT00,SY08,KT08}.
It is tempting to assume that $\nu=3$ 
is just the number of the $SL(2,\Bbb{R})$ zero modes,
whose contribution has gotten regularized by nonvanishing $\del$.
This issue will be addressed elsewhere.

\vspace*{-4mm}
\begin{acknowledgments}
\vspace*{-2mm}
We are grateful to Emil Akhmedov, Pawel Caputa, 
Charlotte Kristjansen, Andrey Mironov,
and Gordon Semenoff for useful discussions.
J.A. thanks FNU, the Danish Research Council for 
Independent Research,  for financial support via the project ``Quantum Gravity
and the Role of Black Holes''. 
Y.M. thanks the NBI High Energy Theory group for hospitality and 
financial support. 
\end{acknowledgments}

\appendix

\vspace*{2mm}
\section{Momentum-space analysis\label{appA}}

We can handle \eq{integ} by 
introducing the one-dimensional Fourier transformation 
\be
\beta(p)=\int \d s\, \beta(s) \e^{\i p s},
\ee
which is of the type of the mode expansion~\rf{modes}.
Noting that 
\begin{widetext}
\be
 D(p) \equiv 
\int \d s\,\e^{\i p s} \left[\frac{1}{(s_1-s_2)^4}  \right]_{\rm reg} = 
-\frac{\pi \e^{-2|p|\ym}}{2\ym^3} (1+ |p|\ym)(1+ |p|\ym + |p|^2\ym^2),
\ee
we obtain
\be
S_{\rm reg} =\frac1\pi \int \frac{\d p}{2\pi}\frac{\d q}{2\pi} 
\, \beta(\frac q2 +p) \beta(\frac q2-p)\int \d s
\,\e^{-\i q s} \left(D(0)-D(p)\right).
\label{Uuufff}
\ee
\end{widetext}
The abovementioned  $SL(2,\Bbb{R})$ symmetry of the right-hand side 
is manifest 
because the Fourier-transformed 
zero modes~\rf{sl2t} are
\be
\beta_1(p)=\delta(p),\quad
\beta_2(p)=\delta^\prime(p),\quad
\beta_3(p)=\delta^{\prime\prime}(p).
\ee

A subtlety with the expression on the right-hand side of \eq{Uuufff}
is that $D(p)$ depends on $s$, as $\ym$ does
according to \eq{ymvsrm} with $t=s$. Otherwise it would simply involve 
$\delta(q)$ after integrating over $s$.
Nevertheless, we do not expect a cancellation 
on the right-hand side of \eq{Uuufff} for generic values of $p$.
We therefore evaluate 
\be
S_{\rm reg} \sim \left(\frac{R}{\del} \right)^3
\ee
which corresponds to $\nu=3$.

For constant $\ym$ we can rigorously obtain this behavior by evaluating 
the path integral over $\beta(p)$, whose contribution to the effective action
reads
\be
\int \d p\, \ln\left(D(p)-D(0) \right) = \frac 32 \ln \ym +{\cal O}(1),
\ee
where the $\zeta$-function regularization has been used again. 
The use of the  $\zeta$-function regularization
can be justified by a conformal mapping of UHP onto
a unit disk, whose boundary is a circle parametrized by 
$\sigma =2 \arctan s$. Then the mode expansion
goes in $\exp(\i p\sigma)$ with integer $p$.

It is now clear that the same consideration as for constant $\ym$ 
is applicable for our case
of \eq{ymvsrm} as well, because the $(R/\del)^3$ factor factorizes in
$(D(p)-D(0))$ on the right-hand side of \eq{Uuufff}.

Finally we mention that $D(p)-D(0)=\pi |p|^3/6+ {\cal O}(\ym^2)$
as $p\ll1/\ym$.
Such an emergence of $|p|^3$ for $AdS$ instead of $|p|$ as in
flat space [the latter dependence stems from \eq{oDou}] was first emphasized 
in Ref.~\cite{Rych}.


\end{document}